\documentclass[prd,twocolumn,amsmath,amssymbm,prd]{revtex4}
\usepackage{epsfig}
\usepackage{graphicx}
\usepackage{amssymb}

\begin{document}
\title{Scattering by a long-range potential}
\author{Shahar Hod}
\address{The Ruppin Academic Center, Emeq Hefer 40250, Israel}
\address{ }
\address{The Hadassah Institute, Jerusalem 91010, Israel}
\date{\today}

\begin{abstract}
\ \ \ The phenomenon of wave tails has attracted much attention over
the years from both physicists and mathematicians. However, our
understanding of this fascinating phenomenon is not complete yet. In
particular, most former studies of the tail phenomenon have focused
on scattering potentials which approach zero asymptotically
($x\to\infty$) faster than $x^{-2}$. It is well-known that for these
(rapidly decaying) scattering potentials the late-time tails are
determined by the first Born approximation and are therefore {\it
linear} in the amplitudes of the scattering potentials (there are,
however, some exceptional cases in which the first Born
approximation vanishes and one has to consider higher orders of the
scattering problem). In the present study we analyze in detail the
late-time dynamics of the Klein-Gordon wave equation with a ({\it
slowly} decaying) Coulomb-like scattering potential:
$V(x\to\infty)=\alpha/x$. In particular, we write down an explicit
solution (that is, an exact analytic solution which is not based on
the first Born approximation) for this scattering problem. It is
found that the asymptotic ($t\to\infty$) late-time behavior of the
fields depends {\it non}-linearly on the amplitude $\alpha$ of the
scattering potential. This non-linear dependence on the amplitude of
the scattering potential reflects the fact that the late-time
dynamics associated with this slowly decaying scattering potential
is dominated by {\it multiple} scattering from asymptotically far
regions.
\end{abstract}
\maketitle


\section{Introduction}

It is well-known that waves and fields propagating under the
influence of a scattering potential do not cut off sharply after the
passage of the primary pulse. Instead, in the presence of a
scattering potential, propagating waves tend to die off gently,
leaving behind ``tails" which decay to zero only asymptotically. The
precise description of these late-time tails is an important subject
in the scattering theory of waves and fields.

Late-time decaying tails arise naturally in a variety of physical
situations. For instance, the seminal work of Price \cite{Price} has
revealed that the late-time dynamics of massless fields propagating
in (curved) black-hole spacetimes are characterized by inverse
power-law decaying tails. The phenomenon of wave tails in curved
spacetimes has since been studied by many researches, see
\cite{Tails} and references therein.

The physically most interesting mechanism for the production of
asymptotically ($t\to\infty$) decaying tails is the backscattering
of the waves by an effective scattering potential at asymptotically
($x\to\infty$) far regions \cite{Price,Thor}. In particular, the
propagation of waves and fields is often governed by a Klein-Gordon
wave equation of the form \cite{Notecc}
\begin{equation}\label{Eq1}
\Big[{{\partial^2}\over{\partial t^2}}-{{\partial^2}\over{\partial
x^2}}+V(x)\Big]\Psi(x,t)=0\  ,
\end{equation}
where $V(x)$ is an effective scattering potential [It is well-known
that the wave equation (\ref{Eq1}) with $V\equiv 0$ describes a
tail-free propagation of the fields].

In a brilliant work Ching et. al. \cite{Ching} provided a simple and
intuitive explanation for the tail phenomena: consider a wave from a
source point $y$ and an observer which is located at a fixed spatial
point $x_{\text{observer}}$. The asymptotic ($t\to\infty$) late-time
tail observed at the point $x_{\text{observer}}$ is a consequence of
the waves first propagating to a distant point $x'\gg
y,x_{\text{observer}}$, being backscattered by the scattering
potential $V(x')$ at $x'$, and then returning to the location
$x_{\text{observer}}$ of the observer at a time $t\simeq
(x'-y)+(x'-x_{\text{observer}})\simeq 2x'$. Note that according to
this heuristic picture, the scattering amplitude is expected to be
proportional to ({\it linear} in) $V(x'\simeq t/2)$.

Despite of the numerous works dedicated to the study of
asymptotically decaying wave tails (see \cite{Tails} and references
therein), our understanding of this fascinating phenomenon is not
complete yet. In particular, it is worth noting that most previous
studies of the tail phenomenon have focused on scattering potentials
which approach zero asymptotically ($x\to\infty$) {\it faster} than
$x^{-2}$. It was shown in \cite{Ching,Hod1} that for these
(rapidly-decaying) scattering potentials the late-time tails can be
determined by the {\it first} Born approximation. Accordingly, the
leading-order late-time tails associated with these rapidly decaying
scattering potentials are {\it linear} in the amplitudes of the
potentials \cite{Ching,Hod1,Noteexc}.

Before proceeding further, it is important to mention that there are
some exceptional cases in which the first Born approximation
vanishes and one has to consider higher orders of the scattering
problem \cite{Ching}. That happens for a 3+1  dimensions scattering
potential of the form $V(x)\sim x^{-\alpha}$, when $\alpha$ is an
odd integer in the range $0\leq\alpha-3<2l$ and $l$ is the multipole
index of the mode. This exceptional behavior also happens for the
spherically symmetric mode in odd $2l+3\geq5$ spatial dimensions.
(In fact, it can be shown that the wave equation for the $l$-th
multipole in 3+1 dimensions and the spherically symmetric wave
equation in odd $d=2l+3$ spatial dimensions are two faces of the
{\it same} mathematical problem.)

It is worth emphasizing that much less is known about the tail
phenomenon associated with {\it slowly} decaying scattering
potentials -- scattering potentials which approach zero
asymptotically ($x\to\infty$) {\it slower} than $x^{-2}$. In this
respect it is worth mentioning the work of Ching et. al.
\cite{Ching}, who analyzed the particular case of an inverse square
law scattering (repulsive) potential of the form
\begin{equation}\label{Eq2}
V(x)={{\nu(\nu+1)}\over{x^2}}\  ,
\end{equation}
with $\nu>0$. This scattering potential is actually the boundary
between the well-studied family of {\it rapidly}-decaying potentials
(scattering potentials which approach zero asymptotically faster
than $x^{-2}$) \cite{Ching,Hod1} and the poorly-explored family of
{\it slowly}-decaying potentials (scattering potentials which
approach zero asymptotically slower than $x^{-2}$).

It was found in \cite{Ching} that the `marginally'-decaying
\cite{Notemar} scattering potential (\ref{Eq2}) produces a late-time
tail of the form \cite{Notenu}
\begin{equation}\label{Eq3}
\Psi(t\gg x)\sim A(\nu)t^{-(2\nu+2)}\  .
\end{equation}
The temporal dependence (\ref{Eq3}) of the late-time tail is
obviously not linear in the scattering potential (\ref{Eq2}) [In
particular, the amplitude $A(\nu)$ of the tail (\ref{Eq3}) was found
in \cite{Ching} to depend non-linearly on the amplitude $\nu(\nu+1)$
of the scattering potential (\ref{Eq2})]. This implies that, for
this marginally-decaying scattering potential, the first Born
approximation {\it fails} to describe the correct late-time behavior
of the fields.

To the best of our knowledge, there are no results in the literature
for the tail phenomenon associated with slowly-decaying scattering
potentials -- long-range potentials that approach zero
asymptotically ($x\to\infty$) {\it slower} than $x^{-2}$.

The main goal of the present study is to extend our knowledge about
the tail phenomenon for {\it slowly} decaying scattering potentials.
As explained above, the results of \cite{Ching} for the
marginally-decaying case (\ref{Eq2}) indicate that the first Born
approximation may fail to describe the correct late-time tails
associated with slowly decaying scattering potentials. Thus, the
analytical approximations used in \cite{Ching,Hod1} to analyze the
late-time tails associated with rapidly-decaying scattering
potentials \cite{Notetool} may not be valid in the regime of the
slowly-decaying scattering potentials.

One is therefore forced to go beyond the first Born approximation
when studying the late-time wave dynamics associated with slowly
decaying scattering potentials. This fact prevents one from reaching
general conclusions about these slowly-decaying late-time tails
\cite{Notegen}. Instead, we are forced to solve explicitly ({\it
case-by-case}) the Klein-Gordon wave equation (\ref{Eq1}),
specifying in each case the explicit form of the slowly-decaying
scattering potential $V(x)$.

In the present study we shall analyze the late-time dynamics of the
Klein-Gordon wave equation (\ref{Eq1}) with a slowly-decaying
Coulomb-like scattering potential
\begin{equation}\label{Eq4}
V(x)={{\alpha}\over{x}}\  ,
\end{equation}
where $x\in [0,\infty]$. Here $\alpha>0$ is the amplitude of the
scattering potential \cite{Notealp}. Fortunately, as we shall show
below, one can write down an {\it explicit} solution (that is, an
exact analytic solution which is not based on the first Born
approximation) for the Klein-Gordon wave equation (\ref{Eq1}) with
the Coulomb-like scattering potential (\ref{Eq4}). This fact will
allow us to study analytically the asymptotic late-time dynamics of
the wave fields in the presence of this slowly-decaying scattering
potential.

\section{Formalism}

The temporal evolution of a wave-field whose dynamics is governed by
the Klein-Gordon wave equation (\ref{Eq1}) is given by
\cite{Ching,Mor,Leaver,Ander}
\begin{equation}\label{Eq5}
\Psi(x,t>0)=\int\big[G(x,y;t)\Psi_t(y,0)+G_t(x,y;t)\Psi(y,0)\big]dy\
,
\end{equation}
where the (retarded) Green's function $G(x,y;t)$ satisfies the
relation
\begin{equation}\label{Eq6}
\Big[{{\partial^2}\over{\partial t^2}}-{{\partial^2}\over{\partial
x^2}}+V(x)\Big]G(x,y;t)=\delta(t)\delta(x-y)\  .
\end{equation}
The causality requirement dictates the initial condition
$G(x,y;t<0)=0$. The problem now becomes to find the explicit form of
the Green's function $G(x,y,;t)$; the temporal evolution of the
fields for any initial data can then be obtained by performing the
spatial integration in the characteristic equation (\ref{Eq5}).

In order to find the Green's function $G(x,y;t)$ we first use the
Fourier transform
\begin{equation}\label{Eq7}
\bar G(x,y;\omega)=\int_{0^-}^{\infty} G(x,y;t)e^{\i\omega t}dt\  .
\end{equation}
The Fourier transform $\bar G(x,y;\omega)$ satisfies the
Schr\"odinger-like wave equation \cite{Ching,Mor,Leaver,Ander}
\begin{equation}\label{Eq8}
\Big[{{d^2}\over{dx^2}}+\omega^2-V(x)\Big]\bar
G(x,y;\omega)=-\delta(x-y)\
\end{equation}
and is analytic in the upper half $\omega$-plane. The time-dependent
Green's function, $G(x,y;t)$, can be obtained by the inversion
integral \cite{Ching,Mor,Leaver,Ander}
\begin{equation}\label{Eq9}
G(x,y;t)={{1}\over{2\pi}}\int_{-\infty+ic}^{\infty+ic} \bar
G(x,y;\omega)e^{-\i\omega t}d\omega\ ,
\end{equation}
where $c$ is some positive constant.

The Green's function $\bar G(x,y,\omega)$ can be expressed in terms
of two linearly independent solutions of the homogeneous
Schr\"odinger-like wave equation \cite{Ching,Mor,Leaver,Ander}
\begin{equation}\label{Eq10}
\Big[{{d^2}\over{dx^2}}+\omega^2-V(x)\Big]\bar\psi_i(x;\omega)=0\ \
\ ; \ \ \ i=1,2\  .
\end{equation}
The two basic solutions $\{\bar\psi_1,\bar\psi_2\}$ which are
required in order to build the Green's function $\bar G(x,y,\omega)$
are defined by the two boundary conditions
\cite{Ching,Mor,Leaver,Ander}: the first solution
$\bar\psi_1(x;\omega)$ corresponds to a vanishing field at $x=0$
where the scattering potential is infinitely repulsive,
whereas the second solution $\bar\psi_2(x;\omega)$ corresponds to
purely outgoing waves at $x\to \infty$ [see Eqs. (\ref{Eq19}) and
(\ref{Eq20}) below].

Using the two characteristic solutions $\{\bar\psi_1,\bar\psi_2\}$,
the Green's function $\bar G(x,y;\omega)$ can be written in the form
\cite{Ching,Mor,Leaver,Ander}
\begin{equation}\label{Eq11}
\bar G(x,y;\omega)=-{{1}\over{W(\omega)}}\times
\begin{cases}
\bar\psi_1(x;\omega)\bar\psi_2(y;\omega)
& \text{for}\ \ \ x<y\ ; \\
\bar\psi_1(y;\omega)\bar\psi_2(x;\omega) & \text{for}\ \ \ x>y\  ,
\end{cases}
\end{equation}
where
\begin{equation}\label{Eq12}
W(\omega)=W(\bar\psi_1,\bar\psi_2)=\bar\psi_1\bar\psi_{2,x}-\bar\psi_{1,x}\bar\psi_2\
\end{equation}
is the ($x$-independent) Wronskian of the system.

It proofs useful to bend the integration contour in Eq. (\ref{Eq9})
into the lower-half of the complex $\omega$-plane
\cite{Ching,Mor,Leaver,Ander}. It was shown by Leaver \cite{Leaver}
that in the complex-frequency picture the late-time behavior of the
field (the asymptotic tail) can be associated with the existence of
a branch cut in the Green's function $\bar G(x,y;\omega)$ of the
Klein-Gordon wave equation [see, in particular, Eq. (\ref{Eq25})
below]. This cut is usually placed along the negative imaginary
$\omega$-axis. The asymptotic late-time tail arises from the
integral of $\bar G(x,y;\omega)$ around the branch cut
\cite{Leaver}. In particular, taking cognizance of Eqs. (\ref{Eq9})
and (\ref{Eq11}) one finds that the branch cut contribution to the
Green's function $G(x,y;t)$ is given by
\cite{Ching,Leaver,Ander,Noteobser}
\begin{eqnarray}\label{Eq13}
G^C(x,y;t)={{1}\over{2\pi}}\int_{0}^{-i\infty}\Big[{{\bar\psi_1(y;\omega
e^{2\pi i})\bar\psi_2(x;\omega e^{2\pi i})}\over{W(\omega e^{2\pi
i})}}\nonumber
\\ -{{\bar\psi_1(y;\omega)\bar\psi_2(x;\omega)}\over{W(\omega)}}\Big]e^{-i\omega
t}d\omega\  .
\end{eqnarray}

\section{The explicit (analytic) solution}

We shall now evaluate the integral (\ref{Eq13}) for the branch cut
contribution to the Green's function $G(x,y;t)$. Substituting the
slowly-decaying scattering potential (\ref{Eq4}) into the
characteristic equation (\ref{Eq10}), one obtains the
Schr\"odinger-like wave equation
\begin{equation}\label{Eq14}
\Big({{d^2}\over{dx^2}}+\omega^2-{{\alpha}\over{x}}\Big)\bar\psi(x;\omega)=0\
.
\end{equation}
Defining
\begin{equation}\label{Eq15}
z\equiv -2i\omega x\  ,
\end{equation}
we can write Eq. (\ref{Eq14}) in the form
\begin{equation}\label{Eq16} \Big({{d^2}\over{dz^2}}-{1\over
4}-{{i\alpha/2\omega}\over{z}}\Big)\bar\psi(z)=0\ .
\end{equation}
Equation (\ref{Eq16}) is the familiar Whittaker differential
equation \cite{Abr}. Its two basic solutions which are required in
order to build the Green's function are (see Eqs. 13.1.32 and
13.1.33 of \cite{Abr})
\begin{equation}\label{Eq17}
\bar\psi_1(z)=Aze^{-{1\over 2}z} M(1+i\alpha/2\omega,2,z)\
\end{equation}
and
\begin{equation}\label{Eq18}
\bar\psi_2(z)=Bze^{-{1\over 2}z} U(1+i\alpha/2\omega,2,z)\ ,
\end{equation}
where $M(a,b,z)$ and $U(a,b,z)$ are the confluent hypergeometric
functions \cite{Abr}, and $A$ and $B$ are normalization constants.

Using equations 13.5.5 and 13.1.8 of \cite{Abr}, one finds the
asymptotic behaviors
\begin{equation}\label{Eq19}
\bar\psi_1(x;\omega)\sim x\ \ \ \text{as}\ \ \ x\to 0\  ,
\end{equation}
and
\begin{equation}\label{Eq20}
\bar\psi_2(x;\omega)\sim x^{-i\alpha/2\omega}e^{i\omega x}\ \ \
\text{as}\ \ \ x\to \infty\  .
\end{equation}
Thus, the characteristic solution $\bar\psi_1(x;\omega)$ describes a
vanishing field at $x=0$ where the scattering potential is
infinitely repulsive,
whereas the characteristic solution $\bar\psi_2(x;\omega)$ describes
outgoing waves at $x\to \infty$.

In addition, using Eq. 13.1.22 of \cite{Abr}, one finds that the
$x$-independent Wronskian $W(\bar\psi_1,\bar\psi_2)$ of the system
is given by
\begin{equation}\label{Eq21}
W(\omega)={{2i\omega AB}\over{\Gamma(1+i\alpha/2\omega)}}\  .
\end{equation}

Note that $M(a,b,z)$ is a single-valued function whereas $U(a,b,z)$
is a many-valued function \cite{Abr}. In particular, from Eqs.
(\ref{Eq17}) and (\ref{Eq18}) one finds
\begin{equation}\label{Eq22}
\bar\psi_1(ze^{2\pi i})=\bar\psi_1(z)\  .
\end{equation}
and (see Eq. 13.1.6 of \cite{Abr})
\begin{equation}\label{Eq23}
\bar\psi_2(ze^{2\pi i})=\bar\psi_2(z)+{B\over A}{{2\pi
i}\over{\Gamma(i\alpha/2\omega)}}\bar\psi_1(z)\ .
\end{equation}
From Eqs. (\ref{Eq22}) and (\ref{Eq23}) one finds the simple
relation
\begin{equation}\label{Eq24}
W(\omega e^{2\pi i})=W(\omega)\  .
\end{equation}
Taking cognizance of Eqs. (\ref{Eq21})-(\ref{Eq24}) one finds
\cite{Noterat}
\begin{equation}\label{Eq25}
{{\bar\psi_1(z'e^{2\pi i})\bar\psi_2(ze^{2\pi i})}\over{W(\omega
e^{2\pi i})}}-{{\bar\psi_1(z')\bar\psi_2(z)}\over{W(\omega)}}=
{{i\pi\alpha}\over{2A^2\omega^2}}\bar\psi_1(z)\bar\psi_1(z')\ .
\end{equation}
Substituting Eq. (\ref{Eq25}) into Eq. (\ref{Eq13}), we find
\begin{equation}\label{Eq26}
G^C(x,y;t)=-{{\alpha}\over{4A^2}}\int_0^{\infty}
{{\bar\psi_1(-2\varpi x)\bar\psi_1(-2\varpi y)}\over{\varpi^2}}
e^{-\varpi t} d\varpi\ .
\end{equation}
for the time-dependent Green's function, where $\varpi\equiv
i\omega$.

It is worth emphasizing that the expression (\ref{Eq26}) for the
branch cut contribution to the Green's function is exact for all
times. We shall now focus on the asymptotic ($t\to\infty$) late-time
behavior of the scattered fields.

\section{The asymptotic late-time tail}

In the asymptotic $t\to\infty$ limit the effective contribution to
the integral in (\ref{Eq26}) comes from $\varpi$-values in the
regime $\varpi=O({1\over t})$. This observation is attributed to the
presence of the rapidly decreasing term $e^{-\varpi t}$ which, in
the $t\to\infty$ asymptotic limit, limits the effective contribution
to the integral (\ref{Eq26}) to small $\varpi$-values in the regime
$0\leq \varpi t\lesssim 1$, that is to small $\varpi$-values with
$\varpi_{\text{max}}=O({1\over t})$.

This mathematical observation has a simple physical interpretation:
it is well-known \cite{Ching,Hod1,Leaver,Ander} that the asymptotic
late-time ($t\to\infty$) dynamics is determined by the
backscattering of the field from asymptotically far ($x\to\infty$)
regions. Thus, the late-time dynamics of the field is dominated by
the {\it low}-frequency contribution to the Green's function because
only low frequencies can be backscattered by the small scattering
potential at spatial infinity [remember that $V(x\to\infty)\to 0$].

Since the late-time asymptotic ($t\gg x,y$) behavior of the field is
dominated by the low-frequency contribution to the Green's function
(frequencies with $\varpi x\ll 1$ and $\varpi y\ll 1$), one may use
Eq. 13.3.1 of \cite{Abr} to write
\begin{equation}\label{Eq27}
\bar\psi_1(-2\varpi x)\simeq -2A\varpi
\sqrt{{x}/{\alpha}}I_1(2\sqrt{\alpha x})\  ,
\end{equation}
where $I_1(z)$ is the modified Bessel function \cite{Abr}.
Substituting (\ref{Eq27}) into (\ref{Eq26}), one obtains
\begin{equation}\label{Eq28}
G^C(x,y;t)=-\sqrt{xy}I_1(2\sqrt{\alpha x})I_1(2\sqrt{\alpha
y})\int_0^{\infty} e^{-\varpi t} d\varpi\ .
\end{equation}
Performing the integration in (\ref{Eq28}), we finally find
\begin{equation}\label{Eq29}
G^C(x,y;t\to\infty)=-{{\sqrt{xy}I_1(2\sqrt{\alpha
x})I_1(2\sqrt{\alpha y})}\over{t}}\
\end{equation}
for the asymptotic late-time behavior of the Green's function
\cite{Notecorr1}. The temporal evolution of the field,
$\Psi(x,t\to\infty)$, can now be obtained by substituting the
expression (\ref{Eq29}) for the time-dependent Green's function into
the characteristic equation (\ref{Eq5}).

We learn from Eq. (\ref{Eq29}) that the asymptotic ($t\gg
x_{\text{observer}}$) behavior of the field is {\it not} linear in
the amplitude $\alpha$ of the Coulomb-like scattering potential
(\ref{Eq4}). Thus, the late-time behavior associated with the slowly
decaying scattering potential (\ref{Eq4}) is determined by {\it
multiple} scattering from asymptotically far regions. This confirms
our previous expectation that, for {\it slowly} decaying scattering
potentials (scattering potentials which approach zero slower than
$x^{-2}$), the first Born approximation fails to describe the
correct late-time behavior of the fields \cite{NoteBor}.

\section{The near-region}

The expression (\ref{Eq29}) for the asymptotic behavior of the
time-dependent Green's function can be simplified if we assume that
the observer is situated in the near-region
\begin{equation}\label{Eq30}
x_{\text{observer}}\ll \alpha^{-1}\  .
\end{equation}
In this regime one may use Eq. 9.6.7 of \cite{Abr} to simplify the
Green's function (\ref{Eq29}):
\begin{equation}\label{Eq31}
G^C(x,y;t\to\infty)=-xy\times {{\alpha}\over{t}}\ \ \ \text{for}\ \
\ x\ll\alpha^{-1}\  .
\end{equation}

We learn from Eq. (\ref{Eq31}) that the asymptotic dynamics of the
field in the region $x_{\text{observer}}\ll\alpha^{-1}$ is
determined by the {\it first} Born approximation. That is, in the
regime (\ref{Eq30}) the leading-order late-time tail is {\it linear}
in the amplitude of the scattering potential:
$\Psi(x,t\to\infty)\propto V(t/2)$ \cite{Notecorr2}.

\bigskip



%
%

\section{Summary}

The time evolution of wave fields governed by the Klein-Gordon wave
equation (\ref{Eq1}) with the (slowly-decaying) Coulomb-like
scattering potential (\ref{Eq4}) was investigated. It was shown that
one can write down an explicit solution (that is, an exact analytic
solution which is not based on the first Born approximation) for
this scattering problem. This fact allowed us to study analytically
the asymptotic late-time dynamics of the fields.

The asymptotic ($t\to\infty$) behavior of the fields was found to
depend {\it non}-linearly on the amplitude $\alpha$ of the
scattering potential [see Eq. (\ref{Eq29})]. This non-linear
dependence on the amplitude of the scattering potential reflects the
fact that the late-time dynamics of the wave equation (\ref{Eq1}) in
the presence of the slowly-decaying scattering potential (\ref{Eq4})
is dominated by {\it multiple} scattering from asymptotically far
regions.

Finally, we have shown that the late-time behavior of the fields in
the near-region $x_{\text{observer}}\ll\alpha^{-1}$ is determined by
the first Born approximation. That is, in this regime the
leading-order late-time tail is {\it linear} in the amplitude
$\alpha$ of the scattering potential.

\bigskip
\noindent
{\bf ACKNOWLEDGMENTS}
\bigskip

This research is supported by the Carmel Science Foundation. I thank
Yael Oren, Arbel M. Ongo and Ayelet B. Lata for helpful discussions.



\begin{thebibliography}{99}
\bibitem{Notecc} We use units in which $c=1$. Thus, both $x$ and $t$
have the dimensions of length.
\bibitem{Noteexc} See, however, \cite{Hod1} for some
modifications to this simple heuristic picture in cases where
$V'(x)$ approaches zero slower than $V(x)/x$ as $x\to\infty$.
\bibitem{Notealp} Note that the amplitude $\alpha$ has the dimensions of
length$^{-1}$.
\bibitem{Noteobser} For simplicity we assume here that the initial field has a considerable
support only inside the observer, that is for $y$-values with $y<x$.
\bibitem{Noterat} Here we have used the relation
$\Gamma(1+i\alpha/2\omega)/\Gamma(i\alpha/2\omega)=i\alpha/2\omega$,
see Eq. 6.1.15 of \cite{Abr}.
\bibitem{Notecorr1} We have not found in the physical (nor in the mathematical)
literature the sub-leading correction terms to Eq. 13.3.1 of
\cite{Abr} for the asymptotic form of $M(a,b,\bar z/a)$ with
$a\to\infty$. However, numerically we find that for $b=2$ the
leading-order correction to the formula 13.3.1 of \cite{Abr} is well
described by a factor $(1+\beta\cdot \bar z/a)$ on the r.h.s of this
equation with $\beta\sim 0.36$. Thus, an improved approximation for
$\bar\psi_1(-2\varpi x)$ can be written as [see Eqs. (\ref{Eq15}),
(\ref{Eq17}), and 13.3.1 of \cite{Abr}] $\bar\psi_1(-2\varpi
x)\simeq -2A\varpi x e^{\varpi x}(\alpha x-2\varpi
x)^{-1/2}I_1(2\sqrt{\alpha x-2\varpi x})(1-2\beta\varpi x)$.
Expanding this expression to second order in $\varpi$ one finds
$\bar\psi_1(-2\varpi x)\simeq -2A\varpi
\sqrt{{x}/{\alpha}}I_1(2\sqrt{\alpha x})[1+\varpi\cdot f(x;\alpha)]$
where $f(x;\alpha)\equiv
x(1-2\beta)-2\sqrt{x/\alpha}I_2(2\sqrt{\alpha x})/I_1(2\sqrt{\alpha
x})$ (Here we have used Eq. 9.6.28 of \cite{Abr}). Thus, the
leading-order correction to Eq. (\ref{Eq28}) is described by the
factor $1+\varpi[f(x;\alpha)+f(y;\alpha)]$ on the r.h.s of this
eqution. Performing the integration in (\ref{Eq28}), one finds that
the leading-order correction to the Green's function (\ref{Eq29}) is
described by the factor $1+[f(x;\alpha)+f(y;\alpha)]/t$ on the r.h.s
of this equation. As expected, this correction factor approaches $1$
in the asymptotic late-time limit $t\to\infty$.
\bibitem{Notecorr2} Using Eq. 9.6.10 of \cite{Abr}, one finds that
the leading-order correction to Eq. (\ref{Eq31}) is given by a
factor $1+\alpha^2(x^2+y^2)/2$ on the r.h.s of this equation. Note
that this correction factor is non-linear in the amplitude $\alpha$
of the scattering potential and, as expected, it approaches $1$ in
the regime $\alpha x_{\text{observer}}\ll1$.
\end{thebibliography}
\end{document}